\documentclass[draft]{agujournal2019}
\usepackage{url} 
\usepackage{lineno}
\usepackage[inline]{trackchanges} %for better track changes. finalnew option will compile document with changes incorporated.
\usepackage{soul}
\usepackage{bm}
\draftfalse

\journalname{arXiv}

\begin{document}

\title{Fault gouge failure induced by fluid injection: Hysteresis, delay and shear-strengthening.}

\authors{Pritom Sarma\affil{1,2}, Einat Aharonov\affil{1,4}, Renaud Toussaint\affil{3,4}, Stanislav Parez\affil{5,6}}

\affiliation{1}{Institute of Earth Sciences, Hebrew University of Jerusalem, Jerusalem, Israel}
\affiliation{2}{Department of Earth and Environmental Sciences, Tulane University, New Orleans, LA, USA}
\affiliation{3}{Université de Strasbourg, CNRS, ENGEES, Institut Terre et Environnement de Strasbourg, UMR7063, Strasbourg, France}
\affiliation{4}{PoreLab, The Njord Centre, Departments of Physics and Geosciences, University of Oslo, Oslo, Norway}
\affiliation{5}{Institute of Chemical Process Fundamentals, Czech Academy of Sciences, Prague, Czech Republic}
\affiliation{6}{Faculty of Science, Jan Evangelista Purkyně University, Ústí nad Labem, Czech Republic}

\correspondingauthor{Pritom Sarma}{pritom.sarma@mail.huji.ac.il}

\begin{keypoints}
\item	Fluid injection into granular fault gouge, shows hysteresis in friction and shear-rate strengthening.
\item	Fluid presence delays failure, associated with pre-failure dilative strain accumulation.

\item   Dilation and fluid diffusion act as competing mechanisms towards the onset of slip.
\end{keypoints}

\begin{abstract}

Natural faults often contain a fluid-saturated, granular fault-gouge layer, whose failure and sliding processes play a central role in earthquake dynamics. Using a two-dimensional discrete element model coupled with fluid dynamics, we simulate a fluid-saturated granular layer, where fluid pressure is incrementally raised. At a critical fluid pressure level, the layer fails and begins to accelerate. When we gradually reduce fluid pressure, a distinct behavior emerges: slip-rate decreases linearly until the layer halts at a fluid pressure level below that required to initiate failure. During this pressure cycle the system exhibits (1) velocity-strengthening friction and (2) frictional hysteresis. These behaviors, well established in dry granular media, are shown to extend here to shear of dense fluid-saturated granular layers. Additionally, we observe a delay between fluid pressure increase and failure, associated with pre-failure dilative strain and "dilational-hardening". During the delay period, small, arrested slip events dilate the layer in preparation for full-scale failure. Our findings may explain (i) fault motion that continues even after fluid pressure returns to pre-injection levels, and (ii) delayed  failure in fluid-injection experiments, and (iii) pre-failure arrested slip events observed prior to earthquakes.

\end{abstract}

\section*{Plain Language Summary}
Faults in the earth’s crust are often made up of pulverized granular material called gouge, and are usually saturated with fluid. The onset and cessation of slip in fault gouges are crucial to understand how earthquakes are hosted on these faults. We, using a coupled fluid-granular deformation model, simulated a granular layer representative of a fault gouge, exposed to an increase of fluid injection pressure beyond a critical level such that the layer starts sliding. Then the pressure is decreased to the starting level and the sliding slows down, but does not cease until the pressure is well below the failure value, reflecting a hysteresis in friction. Our granular layer also showed increased friction with sliding speed, which along with the hysteresis in friction, is in line with observations in dry granular materials. Characteristic to the presence of fluid, we observe a delay between injection and failure, linked to dilation that occurs primarily via small precursory slip events that are quickly arrested. These findings may help explain continued fault slip after shut-in of fluid injection, delayed onset of rupture and precursory slow slip events prior to major earthquakes.

\section{Introduction}
Active geological faults, which produce earthquakes in the Earth's brittle crust, frequently contain a layer of granular material (GM), referred to as the "fault gouge" layer, which forms by wear of the wallrock during sliding. In addition, faults are usually saturated with fluids. The saturated gouge's mechanical response, composition, and evolution, control the fault zone’s mechanical strength, frictional stability, and seismic slip risk. Since friction controls the stability of faults and their propensity for earthquakes (e.g. \citeA{marone1998laboratory, nielsen2000influence, kanamori2004physics, scholz2019mechanics}), it is crucial to comprehend the behavior of wet and dry gouge-filled fault zones. 

One of the major controls on friction is its velocity dependency \cite{dieterich1978time,ruina1983slip, matsu1992slip, marone1998laboratory, abercrombie2005can, dunham2007conditions, scholz2019mechanics, bhattacharya2022evolution}. Friction that increases with shear rate (velocity-strengthening) is generally thought to produce stable slip or creep. On the other hand, velocity-weakening behavior is considered a requirement for earthquake initiation on faults \cite{rice1983stability, kilgore1993velocity, dieterich1996implications, dieterich1996imaging, baumberger1999physical, hawthorne2013laterally}. Strain rate dependence of friction controls both granular gouge and bare surfaces behaviors \cite{marone1998laboratory}. Shearing GM show variable  velocity dependence: At higher normal stresses ($\sigma_n>1$MPa), which are relevant for natural geological faults, GM shearing at steady state generally shows mild velocity weakening till a velocity of 1 cm s$^{-1}$, beyond which friction abruptly drops with velocity, most likely due to thermal effects \cite{di2011fault, aharonov2018physics,aharonov2019brittle}. Under non-steady-state conditions GM exhibit velocity strengthening \cite{marone1990frictional}, until thermal softening kicks in at seismic rates \cite{rubino2022intermittent}. At low confining stresses on the other hand, dry GM present strong velocity-strengthening friction \cite{forterre2008flows}, following the $\mu(I)$ rheology, where the friction coefficient of a shearing granular layer monotonically increases with the inertial number $I=\dot{\gamma} d / \sqrt{\sigma_n'/\rho_s}$, a non-dimensional number defined as the ratio of the timescale of microscopic to macroscopic deformation, relating strain rate $\dot{\gamma}$, effective normal stress $\sigma_n'$, particle size $d$ and density, $\rho_s$ (e.g., \citeA{da2002viscosity, GDRMiDi, daCruz2005, forterre2008flows}). 

In addition to strain-rate dependence, there is also a dependence of friction (in both GM and bare surfaces) on strain. Slip-weakening has been suggested to control the transition to slow slip, and transition from aseismic to seismic fault slip \cite{matsu1992slip, uenishi2003universal, ikari2013slip}.
For GM, a large part of friction dependency on slip arises from dilation \cite{marone1990frictional} and grain rearrangements. This is because sheared dense GM tend to dilate \cite{reynolds1885lvii, mead1925geologic, aharonov1999rigidity, makedonska2011friction}, and by  energy considerations it can be shown that during  dilatancy effective friction is increased \cite{poulos1971stress, frank1965dilatancy, marone1990frictional, marone1991note, makedonska2011friction, parez2021strain}. 

Friction depends not only on strain and strain-rate: experimental and theoretical studies reveal multi-valued frictional forces in sheared granular layers of polydisperse particles during loading and unloading cycle, causing hysteresis. This means that the phase transition from a jammed state to a flow-like state, and back to a jammed state, essentially follows two different stress-strain paths. As a result, the material's loading history affects its stability \cite{nasuno1997friction, daniels2005hysteresis, garcia2008strong, degiuli2017friction}. Frictional hysteresis in GM implies that stress must be considerably lowered below its failure level, in order to halt a granular layer already in motion, and deceleration will have different stress-strain path compared to acceleration. The hysteresis concept comprises the well-known difference between static and dynamic frictions, as well as the pathway to achieve these values, including the strain-rate dependence of dynamic friction.

Pore fluids also have a profound effect on shear strength and stability of sliding. The effect of pore fluid pressure, $P$, on the resisting shear stress, $\tau$, is commonly expressed by the effective stress law \cite{Hubbert1959, terzaghi1948soil, goren2013general, abbasov2024modeling}
\begin{equation}\label{eqn:eff_stress_law}
\tau = \mu \sigma_\mathrm{n}^\prime = \mu(\sigma_\mathrm{n} - P) \,,
\end{equation}
where $\mu$ is the friction coefficient, and $\sigma_\mathrm{n}$ is fault-normal stress. Increasing pore pressure reduces the shear strength by reducing the effective normal stress,  $\sigma_\mathrm{n}^\prime=\sigma_\mathrm{n} - P$. While high pore pressure facilitates fault activation, stability of sliding is related to a variation of strength with slip and slip rate, as stated above. The presence of fluids in GM leads to variations frictional strength with shear-rate, which in the inertial regime follows  the $\mu(J)$ rheology, a similar rate-strengthening constitutive rheology as the $\mu(I)$ rheology,  but here $J$ is the ratio of the timescale of viscous flow to the timescale of macroscopic deformation and $\mu$ depends on both $J$ and the shear-rate \cite{pailha2008initiation, rauter2021compressible, fei2023mu, fei2024frictional}.

The pore-pressure in Eq.~\ref{eqn:eff_stress_law} often varies with time or slip. An important control on pressure  variability arises from 
dilation accompanying the initiation of slip in well-consolidated faults. Such dilation leads to transient depressurization and dilatant hardening \cite{frank1965dilatancy, scholz1973earthquake, segall1995dilatancy, Parez2023b}, which contributes to slip stabilization \cite{scholz1988mechanisms}. Many laboratory tests and field studies nevertheless demonstrate that fluids can trigger both aseismic and seismic slip \cite{ellsworth2013injection, Niemeijer2014, guglielmi2015seismicity, scuderi2016role, keranen2018induced, noel2019time, schultz2020hydraulic}.

Large subsurface fluid volume injection is increasingly common across a range of economic (enhanced oil recovery, geothermal energy extraction) and remediation ($CO_2$ sequestration, wastewater injection) activities. This fluid injection is often associated with increased seismicity \cite{keranen2014sharp, grigoli2017current, schultz2020hydraulic},  usually ascribed to reduction in fault strength on pre-stressed faults once  diffusing pore-pressure reaches the fault. Additionally, some in-situ experiments, field observations and theoretical works have also suggested that fluid-injection might, instead, primarily induce aseismic fault slip \cite{cornet1997seismic, cornet2016seismic, guglielmi2015seismicity, bhattacharya2019fluid}. Although, the interaction of pore-pressure with GM, specifically with respect to slip, slip-rate dependence and the onset of failure of GM in the presence of fluids, are not well understood. In particular, the question of hysteresis in fluid-saturated GM, and its affect on the frictional stability of granular faults under fluid injection, has been largely unexplored. 

This work uses a coupled fluid-DEM model to simulate a water-saturated gouge-filled fault zone under shear, focusing on the onset of sliding upon pore-pressure increase in saturated fault-gouge layers. We then study the response of the layer to a subsequent decrease in fluid pressure, until motion arrests. We follow this pressure increase-decrease protocol in order to compare the well establish hysteresis and strain-rate in dry GM, to granular layers subjected fluid injection. 
Our model allows us to investigate the coupled response of GM plus pore fluid, without imposing continuum equations describing grain deformation, as these are not well established, especially when the GM transitions between static and flowing states \cite{aharonov1999rigidity, aharonov2004stick, GDRMiDi, daCruz2005, forterre2008flows}, as occurs here. 

The Discrete Element Model (DEM) \cite{Cundall1979}, which we use to simulate the solid grain phase, is a discrete numerical technique that predicts the behavior of granular aggregates by calculating interactions between a large number of grains \cite{aharonov1999rigidity, aharonov2002shear, aharonov2004stick, morgan2004particle, einav2007breakage, ben2010role, tordesillas2011structural,  ferdowsi2015acoustically, ferdowsi2020granular, parez2021strain, arran2024simulated}. Since continuum equations to predict GM deformation are  lacking or insufficient, DEM is instead often used to replicate phenomena and devise theories for the physics of granular fault slip, probing the complicated behavior seen in the lab or in nature. For instance, DEM was used to illuminate the rigidity phase-transition in GM \cite{aharonov1999rigidity}, stick-slip instabilities \cite{aharonov2002shear, aharonov2004stick, lu2007shear, ciamarra2010unjamming, mair2007nature,ferdowsi2013microslips}, seismic emissions during granular avalanches \cite{arran2024simulated} and localization in dry GM \cite{morgan1999numerical, aharonov2013localization,parez2021strain}.

A recent advance in physical modeling capabilities allows modeling DEM coupled with fluid-flow. This has enabled exploration of coupled fluid-solid instabilities (e.g. \citeA{mcnamara2000grains, dorostkar2017role}), insights into the physics of liquefaction \cite{goren2011mechanical,ben2020compaction,ben2023drainage}, shear localization \cite{ yang2018two,parez2021strain}, fault network evolution during fluid injection \cite{ghani2013dynamic, ghani2015dynamics}, fluidization \cite{jajcevic2013large}, and shearing  ice till \cite{kasmalkar2021shear}. 
 Here the coupled fluid-DEM model is utilized to explore the physics controlling onset and cessation of motion in response to gradual pore pressure changes. Results are compared to simulations of dry GM and to a theoretical analysis.

\section{Methods}\label{simulation}
We simulate the grain-level micromechanical interaction within a granular fault gouge in response to increase in traction using the discrete element method (DEM) \cite{Cundall1979}. 
The dynamics of grains is coupled to that of an interstitial fluid by a two-phase numerical model, detailed in Appendix A and in our previous studies  \cite{goren2010pore,goren2011mechanical, ben2020compaction, ben2023drainage, Parez2023b}. In this coupled model DEM uses a generalized force field that includes interaction with a fluid \cite{johnsen2006pattern, vinningland2007experiments, vinningland2007granular}. Fluid dynamics is computed via an Eulerian solver for the pore fluid pressure $P$ \cite{goren2010pore, goren2011mechanical}, solving
\begin{equation}\label{eqn:ppeqn}
\frac{\partial P}{\partial t} - \frac{1}{\beta \phi \eta} \nabla \cdot [k \nabla P] + \frac{1}{\beta \phi} \nabla \cdot u_s = 0 \,,
\end{equation}
in which $\beta$ and $\eta$ are fluid compressibility and viscosity, $\phi$ and $k$ are porosity and permeability of the granular matrix, and $u_s$ is velocity of grains, so that $\nabla \cdot u_s$ is local volumetric strain rate. Negative volumetric strain rate (compaction) leads to fluid pressurization, and, conversely, positive volumetric strain rate (dilation) leads to fluid depressurization.

Table \ref{tab:setup} lists the DEM parameters used in this study. Grain Young's modulus is slightly lower than that of typical gouge material, but on the same order of magnitude. The grain size distribution is obtained from a Gaussian distribution with the mean size and standard deviation both equal to $0.005$~m, and the maximum polydispersity of $\pm 20\%$ is imposed. The size is set larger compared to natural gouge grains \cite{Billi2005} in order to expedite the simulations, so the time scale of grain collisions ($t_\mathrm{0} = d \sqrt{\pi\rho_\mathrm s / 6E}=3.63\times10^{-6}s$) is artificially increased by increasing grain size and decreasing grain Young's modulus, which allows to increase the time step. Our two-phase DEM code was validated under quasi-static loading conditions \cite{goren2011mechanical} by comparing numerical results with experimental data on poro-elastic compression of fluid-filled porous material \cite{nur1971exact}. It has also been validated under dynamic loading conditions by comparison to experiments \cite{ben2020compaction,ben2023drainage,johnsen2006pattern, vinningland2007experiments, vinningland2007granular, vinningland2010size}. For the subsequent part of the work we non-dimensionalized all of the parameters and variables using normalization constants using independent quantities in our study namely grain mean diameter ($d$), grain Young's modulus ($E$) and grain density ($\rho_s$).
\begin{table}[!ht]
\centering
\caption{{\bf Parameter Table} }
\begin{tabular}{|l|l|}
\hline
\multicolumn{1}{|l|}{\bf Parameters} & \multicolumn{1}{|l|}{\bf Value}\\ \hline
Grain Density & $\rho_\mathrm s=2640$ kg m$^{-3}$\\ \hline
Grain Young’s modulus & $E=10^{10}$ Pa \\ \hline
Grain mean diameter& $d=0.01$~m\\ \hline
Grain friction coefficient & $\mu_g=0.5$ \\ \hline
Normal Stress & $\sigma_n=10$ MPa\ \\ \hline
Shear Stress & $\tau=20-30\% \sigma_n$ \ \\ \hline
Injection Pressure & $P_{inj}=20-40\% \sigma_n$ \ \\ \hline
Fluid density & $\rho_\mathrm f=1000$ kg m$^{-3}$\ \\ \hline
Fluid compressibility & $\beta_f=10^{-9}$Pa$^{-1}$\\ \hline
Fluid dynamic viscosity & $\eta=10^{-3}$Pa s\\ \hline
\end{tabular}
\label{tab:setup}
\end{table}

\begin{table}[!ht]

\centering
\caption{{\bf Normalizing Constants} }
\begin{tabular}{|l|l|l|}
\hline
\multicolumn{1}{|l|}{\bf Constant} & \multicolumn{1}{|l|}{\bf Definition}& \multicolumn{1}{|l|}{\bf Value}\\ \hline
Characteristic timescale of grain collision & $t_\mathrm{0} = d\sqrt{\frac{\pi\rho_\mathrm s}{6E}}$  & $3.7 \cdot 10^{-6} s$\\ \hline
Normalizing Velocity & $V^*=\sqrt{\frac{6E}{\pi \rho_s}}$ & $2.7\cdot 10^{3} m/s$\\ \hline
\end{tabular}
\label{tab:normal}
\end{table}

\section{Setup}
\begin{figure}[ht]

\includegraphics[width=1\textwidth]{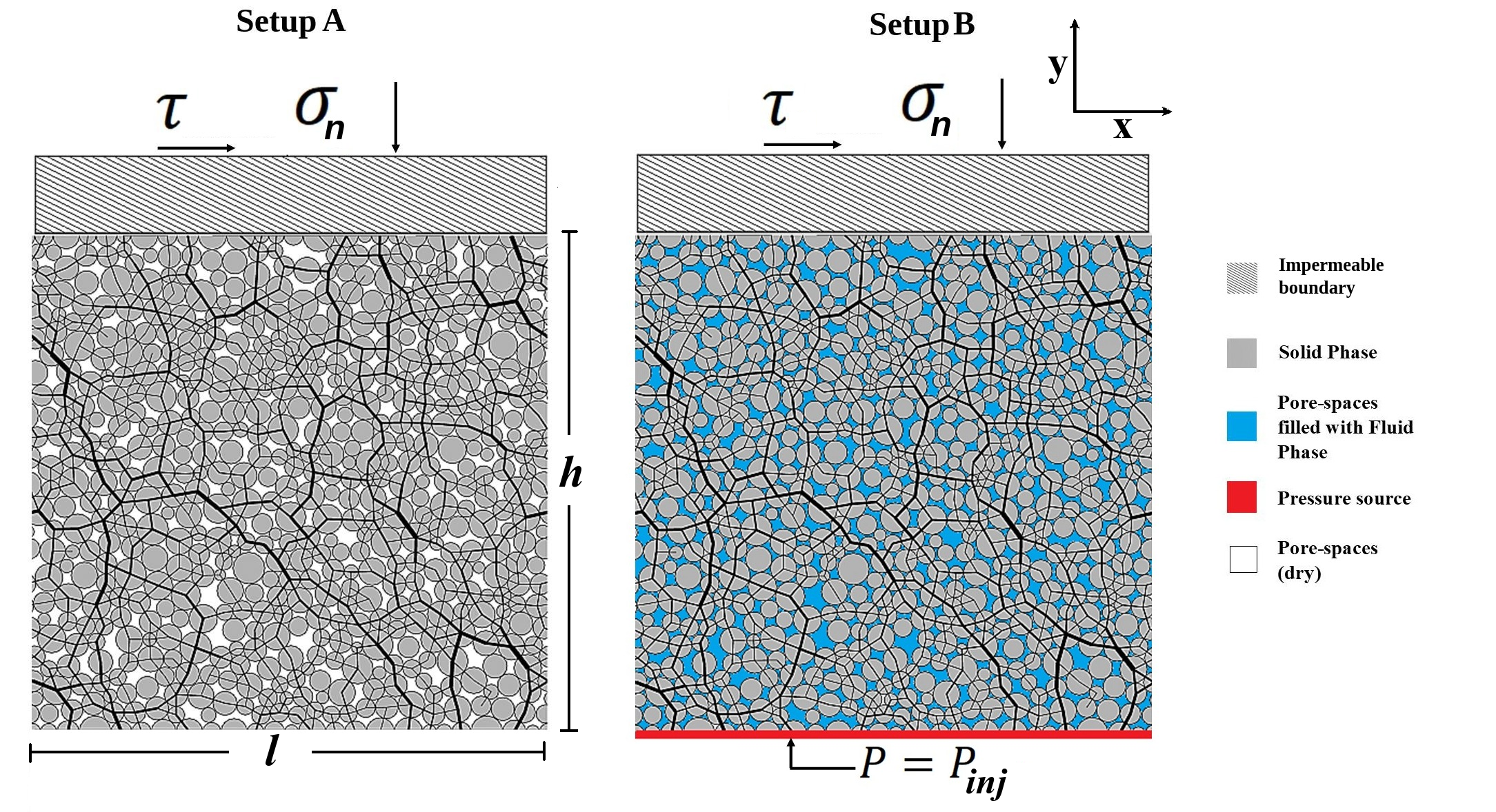}
\caption{Setup of our simulations representing idealized fault geometry. Setup A is a dry layer loaded by incremental shear stress $\tau$ applied on the top wall, while keeping the normal stress $\sigma_n$ fixed. Setup B represents fluid injection into the layer by controlling fluid pressure $P_{inj}$ at the bottom while keeping $\tau=0.2 \sigma_n$ fixed. The bottom boundary is fixed in position and the lateral boundaries are periodic. The thickness of the lines joining the neighboring grain centers is proportional to the normal component of the interparticular force.}
\label{fig:setups}
\end{figure}

The setup of our simulations represents an idealized two-dimensional fault gouge geometry (Fig.~\ref{fig:setups}).
The gouge is conceptualized as a granular layer confined in $z$-direction by two rigid, parallel, rough walls. The walls are constructed from grains glued together into a planar array. Wall grains have the same properties as the interior grains. The bottom wall is static, the top wall is free to move horizontally and vertically. Note that the words top and bottom, horizontal and vertical, are used here with respect to the fault direction: this coincides with the usual terminology for horizontal faults, but the model is identical if the fault considered is not horizontal, and rather with a normal inclined with respect to the direction of gravity. We impose a shear stress $\tau$ and a normal stress $\sigma_n$ on the top wall. Periodic boundary conditions are applied in the $x$ (shear) direction to model a spatially extensive layer. We simulate either a dry system (Setup A) or a wet, fully water-saturated layer (Setup B), where the dry system serves as a control experiment. The fluid-filled layer is in contact with a pressure reservoir along its bottom boundary, at pressure $P_{inj}$. The value of the boundary injection pressure can be kept constant, or modified with time. The top boundary is impermeable. 

The aspect ratio of the layer ($l/h$), where $l$ is the length and $h$ is the thickness of the layer (as shown in Fig.~1), the applied normal stress ($\sigma_n$), the grain friction coefficient ($\mu_{g}$), the applied shear stress ($\tau$), and the injection pressure ($P_{inj}$), are the key parameters of our study, with values ranging between $l/h=0.4-4$, $\sigma_n=10$ MPa, $\mu_{g}=0.5$, $\tau=20-30\% \, \sigma_n$ and $P_{inj}=20-40\% \, \sigma_n$. 

\begin{figure}[htb!]

\centering
\includegraphics[width=1\textwidth]{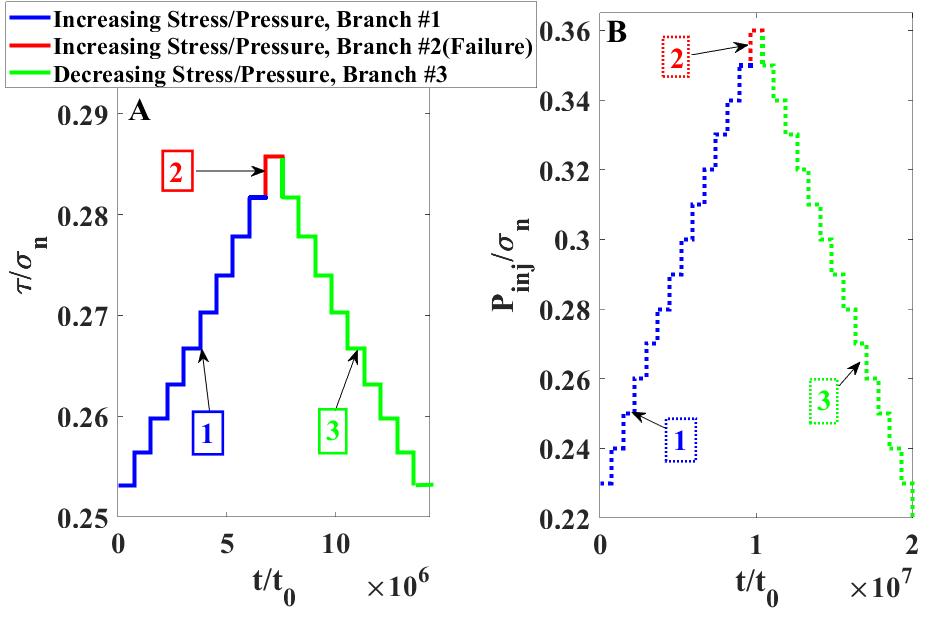}
\caption{Loading path. (A) For dry layers (Setup A in Fig.~\ref{fig:setups}) shear stress is incrementally increased until failure and then decreased. (B) For wet layers (Setup B in Fig.~\ref{fig:setups}) failure is driven by incremental increase in injection pressure until failure, after which it is decreased by steps.}

\label{fig:mod_run}
\end{figure}

Layers are initialized as static, dense granular packings under the applied normal stress and sheared at an initial shear stress, applied on the top wall at 70\% of the shear strength. The initial pore pressure, in the case of the wet systems, is uniform and equal to $P_{inj}=20\% \sigma_n$, which is insufficient to cause failure and sliding of the layer. The layer is then loaded until failure and unloaded until sliding stops. The loading protocol is depicted in Fig.~\ref{fig:mod_run} and uses small step increases either in shear stress (dry system) or $P_{inj}$ (wet system), following the procedure in a benchmark experimental work \cite{cappa2019stabilization}. For a dry layer, shear stress is incrementally increased (blue stairs in Fig.~\ref{fig:mod_run}A) until failure, and then decreased via the same but reverse steps. For a wet layer, the injection pressure $P_{inj}$ is incrementally raised (blue stairs in Fig.~\ref{fig:mod_run}B) until failure and then decreased via a reverse procedure, until the layer comes to rest again.

\section{Results}
\subsection{Hysteresis of strain-rate versus shear stress and pore pressure }

\begin{figure}[ht]

\centering
\includegraphics[width=0.8\textwidth]{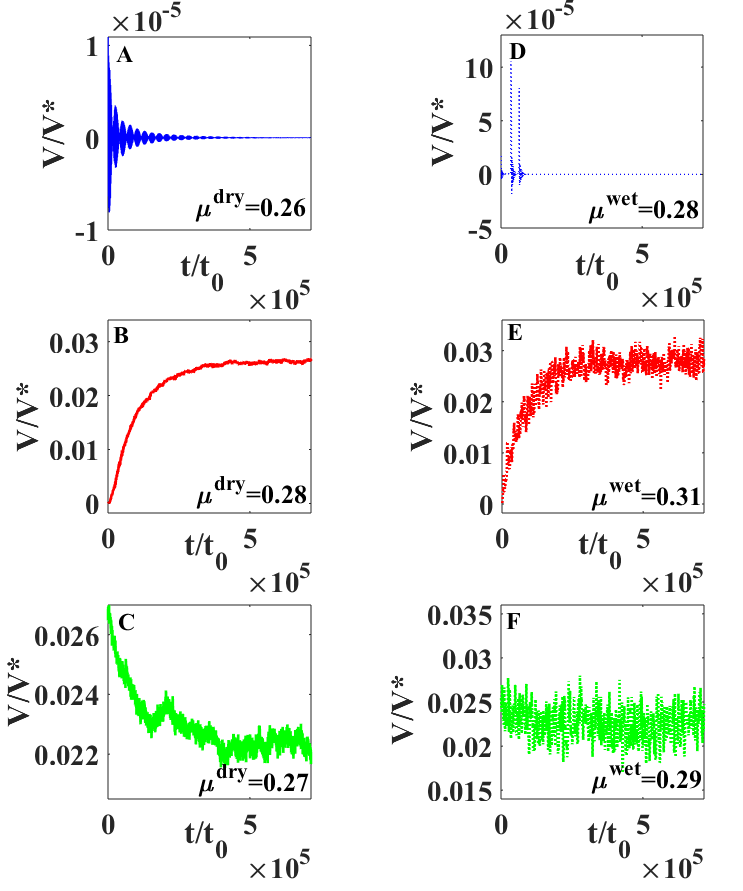}
\caption{Top wall slip-rate evolution following a loading step (the ratio of shear stress to effective normal stress after the loading step is indicated in each plot). Results for dry and wet conditions are displayed on the left and right panels, respectively. Loading branch \#1 (A \& D) is characterized by dampened elastic vibrations or arrested microslips. Branch \#2 (B \& E), which is the failure point, exhibits rapid acceleration and an exponential approach to steady state. Branch \#3 (C \& F), stepping down stress/pressure post-failure, exhibits deceleration.} 

\label{fig:hysterrep}
\end{figure}

\begin{figure}[ht]
\centering
\includegraphics[width=1\textwidth]{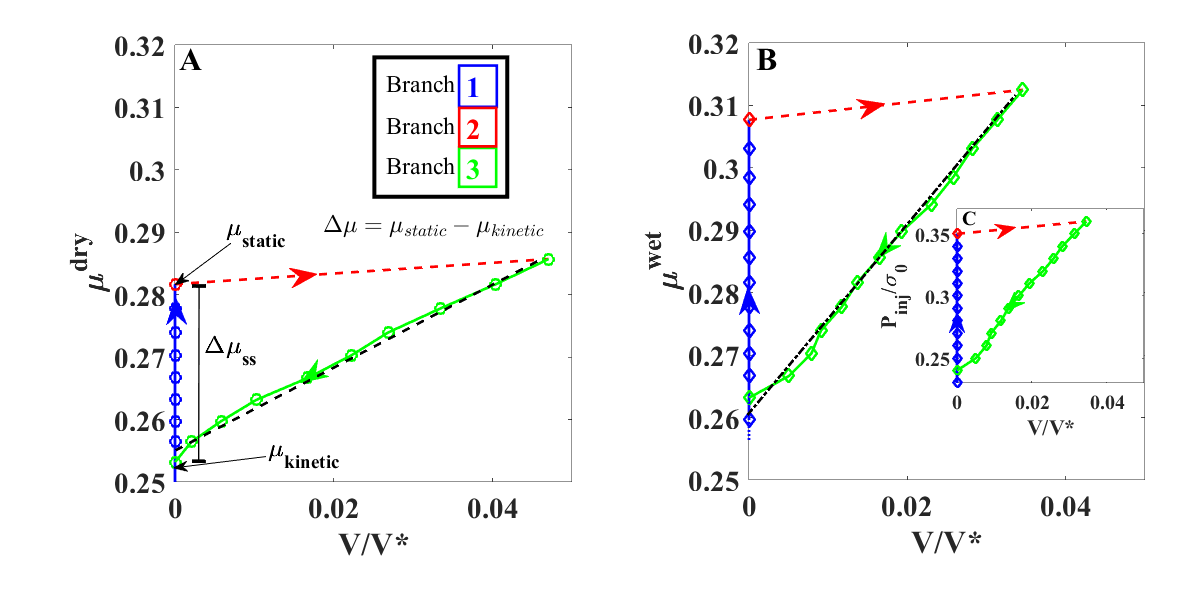}
\caption{Hysteresis loop observed in simulations, for the measured apparent friction coefficient as a function of slip velocity $V$, that is attained after transient acceleration or deceleration following a loading step.
A) Results for dry conditions (setup A in Fig.~\ref{fig:setups}), where the apparent friction coefficient is defined as $\mu^{dry}=\tau/\sigma_n$. B) Results for wet conditions (setup B in Fig.~\ref{fig:setups}), where the apparent friction coefficient is $\mu^{wet}=\tau/(\sigma_n-P_{inj})$. C) Same wet results showing the slip velocity as a function of injection pressure. The color code indicates different branches of the loading protocol (Fig.~\ref{fig:mod_run}). } 
\label{fig:hyster2}
\end{figure}

As shear stress or pore pressure are varied by steps, following loading path A or B in Fig.~\ref{fig:mod_run}, the granular system responds by varying  top-wall and internal velocity, following a hysteresis loop: The loop starts far from failure, when low shear stress/pressure are applied. Each step increase in stress/pressure produces an elastic deformation of the layer and also small irreversible slip events, but the layer finally relaxes, as shown in Figs.~\ref{fig:hysterrep}A (dry) and \ref{fig:hysterrep}D (wet). Fig.~\ref{fig:hysterrep}A displays an elastic response, manifested by dampened shaking of the layer, while Fig.~\ref{fig:hysterrep}D displays micro-slip events, both characterizing the blue branch in Fig.~\ref{fig:hyster2}. This transient response repeats for each increasing shear/injection pressure step, until the condition $\tau=\mu_{static}(\sigma_n - P_{inj})$ is met ($P_{inj}=0$ for the dry system), where $\mu_{static}$ is defined as the friction coefficient at which the layer fails and the top wall starts accelerating. After the initial acceleration, the layer  reaches a steady-state sliding velocity (red branch in Figs.~\ref{fig:hyster2}, \ref{fig:hysterrep}B and \ref{fig:hysterrep}E). The fact that the layer reaches a stable terminal velocity is due to its intrinsic velocity-strengthening behavior, as explained below. We next incrementally reduce the driving shear stress/injection pressure, as depicted by the green branch in Fig.~\ref{fig:mod_run}. After each stress reduction step, we observe the top wall sliding rate slows down to a new, slower steady sliding velocity, as shown in Figs.~\ref{fig:hysterrep}C and \ref{fig:hysterrep}F. Continuing the stress/injection pressure reduction steps shows a linear dependence between stress and resulting steady slip rate (green branch in Fig.~\ref{fig:hyster2}). Finally, when stress/injection pressure is reduced to a low enough level, the top wall stops moving. The stress state associated with cessation of motion satisfies $\tau=\mu_{kinetic}(\sigma_n - P_{inj})$ with $\mu_{kinetic}$ being the friction coefficient when sliding stops, and its value is invariably smaller than $\mu_{static}$. The difference between the two values of friction is found to be around 0.02-0.05. This difference between $\mu_{kinetic}$ and $\mu_{static}$ marks the hysteresis between the friction coefficient and slip velocity. For the wet case  hysteresis is achieved via the change in injection pressure, along with a similar linear dependence between injection pressure and resulting steady slip rate during the down-stepping pressure branch (see Fig.~\ref{fig:hyster2}C).  

\begin{figure}[ht]
\centering
\includegraphics[width=1.0\textwidth]{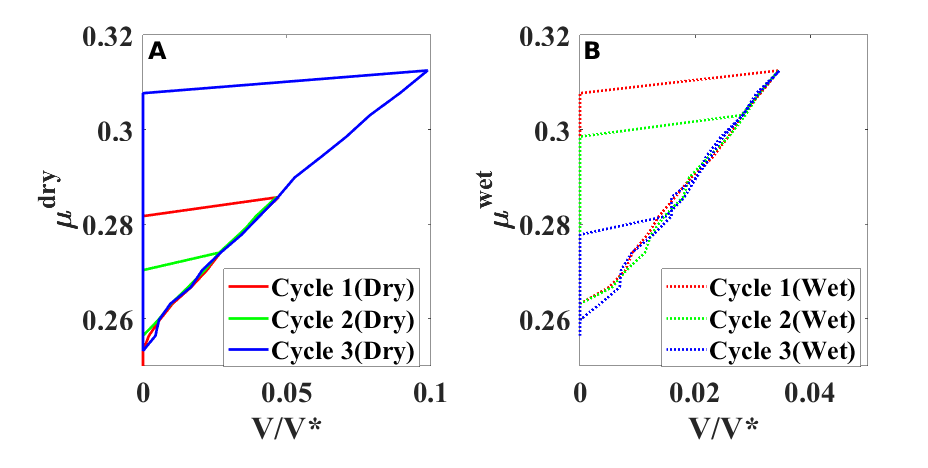}
\caption{The hysteresis loop seen in repeated loading cycles for A) dry and B) wet simulation conditions (Setups defined in Fig.~\ref{fig:setups}). Note that while $\mu_{static}$, the stress at which motion initiates, is highly variable between configurations, $\mu_{kinetic}$, the stress at which motion halts, is constant across different failure events.}
\label{fig:repeatedcycles}
\end{figure}

\begin{figure}[ht]
\centering
\includegraphics[width=0.8\textwidth]{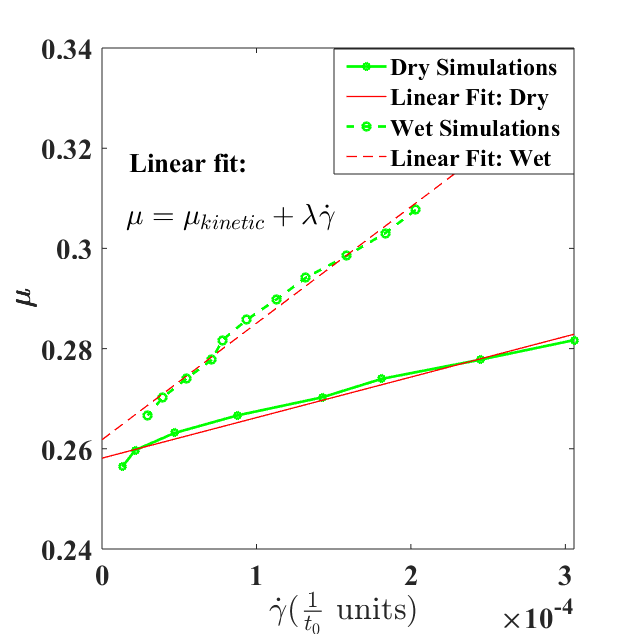}
\caption{Apparent friction coefficient as a function of steady-state shear rate (Eq.~\ref{eqn:stren}) along branch \#3 as in Fig.~\ref{fig:hyster2}, showing rate-strengthening behavior. A linear fit is indicated by solid and dashed red lines for wet (fitting parameter, $\lambda=232.50 t_0$) and dry (fitting parameter, $\lambda=80.94 t_0$) conditions, respectively. }
\label{fig:fit1}
\end{figure}

The loading cycle shown in Fig.~\ref{fig:mod_run} was repeated three times and the three response cycles are shown in Fig~\ref{fig:repeatedcycles}. We find a significant variability in $\mu_{static}$, varying between 0.27-0.31, consistent between dry (control) and wet layers given their scatter. In contrast, $\mu_{kinetic}$ is the same for multiple cycles, and wet layers show slightly higher $\mu_{kinetic}$ compared to dry layers. The observed linear velocity-stress relation during deceleration of the layer (green branch \#3 in Fig.~\ref{fig:hyster2}) is highly reproducible between cycles (but different between wet and dry systems, see Fig.~\ref{fig:fit1}). We have also tried to reverse the loading direction from an arbitrary steady-flow state on branch \#3, i.e. increasing shear stress or injection pressure, and the same velocity-stress curve was recovered. 

\begin{figure}[ht]
\centering
\includegraphics[width=1\textwidth]{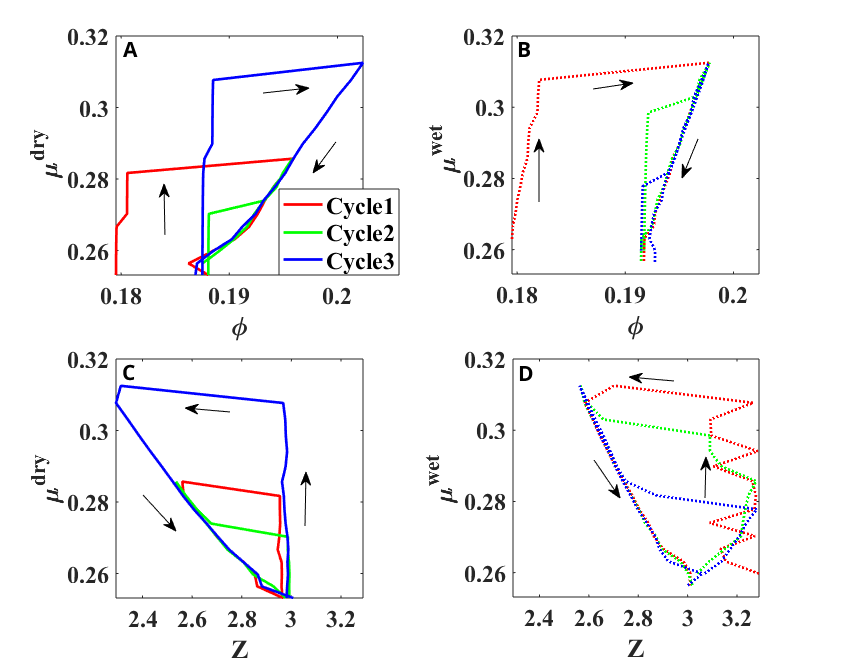}
\caption{Hysteresis in porosity $\phi$ and coordination number $Z$ during 3 loading cycles for dry (A \& C) and wet (B \& D) conditions. The arrows indicate time evolution during each cycle.}
\label{fig:diffpar}
\end{figure}

The main state parameters of our model, the porosity $\phi$ and grain coordination number $Z$  (defined as the average number of touching neighbors per one grain),  exhibit a similar hysteresis as the layers' velocity. Figs.~\ref{fig:diffpar}A and C shows the hysteresis in porosity, $\phi$, across three loading cycles. During the first cycle the layer undergoes a permanent dilation. During each failure dilation occurs, and porosity is reduced as the layer slows down. Figs.~ \ref{fig:diffpar}B and D show the hysteresis in grain coordination number, $Z$. When the layer undergoes failure, the number of touching grains, $Z$, drops, and as the layer decelerates and compacts, $Z$ increases linearly. Comparing dry (control) and wet conditions, wet systems show a smaller difference between porosity in shearing vs static layers, and larger static coordination number.

\begin{figure}[ht]
\centering
\includegraphics[width=0.8\textwidth]{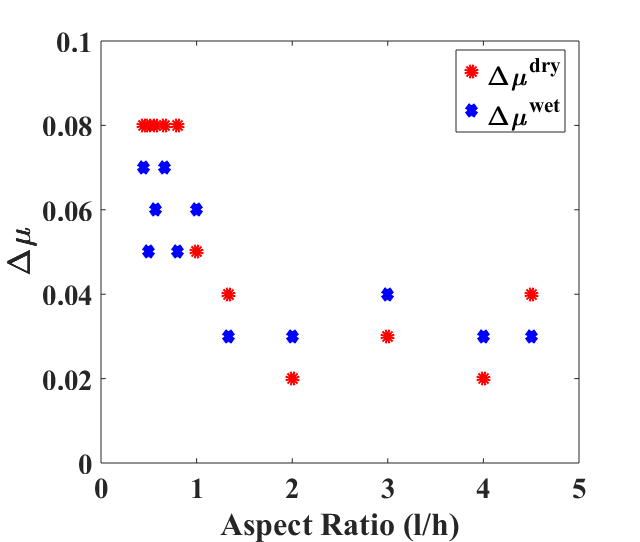}
\caption{Difference between the static and kinetic friction coefficient (amplitude of hysteresis), $\Delta\mu=\mu_{static}-\mu_{kinetic}$, as function of layer aspect ratio $l/h$, for $l$ and $h$ defined in Fig.~ \ref{fig:setups}.}
\label{fig:diff_ig}
\end{figure}

We quantify the amplitude of the hysteresis, and test system size effects, by measuring the drop in friction $\Delta\mu=\mu_{static}-\mu_{kinetic}$, as defined in Fig.~\ref{fig:hyster2}, for different system size ratios. Fig.~\ref{fig:diff_ig} examines how $\Delta\mu$ varies with layer aspect ratio $l/h$. $\Delta\mu$ drops from values of 0.06 - 0.08 for small layer aspect ratios, $l/h<1.5$, to a value of $\Delta\mu \approx 0.03$ for large aspect ratios (elongated  boxes). 

\subsection{Shear-Rate Dependence of Friction}\label{srt}
To examine the rheology controlling the shear, we propose a linear constitutive dependence between velocity and steady-state friction coefficient ($\mu$) of the granular layer, branch \#3 in Fig.~\ref{fig:hyster2}. We therefore adopt a linear constitutive law in which friction increases linearly with shear (strain) rate $\dot{\gamma}=\frac{\partial v}{\partial z}$ at a slope $\lambda$, hereby called the strengthening coefficient,
\begin{equation}
     \mu = \mu_{kinetic} + \lambda \dot{\gamma} \,,
\label{eqn:stren}
\end{equation}
The shear rate $\dot{\gamma}$ can also be expressed via the inertial number of a granular flow $I$ consistent with the $\mu(I)$ rheology. Fig.~\ref{fig:fit1} shows a fit of the branch \#3 from Fig.~\ref{fig:hyster2} to the constitutive law Eq.~\ref{eqn:stren}, using the steady-state shear rate $\dot{\gamma}$ plotted in units of the reciprocal characteristic time of grain collision, $\frac{1}{t_0}$.  Fig.~\ref{fig:fit1} confirms a linear positive trend between stress and strain rate, signifying rate-strengthening behavior, with strengthening coefficient $\lambda$ for the dry and wet simulations equal to $80.94 t_0$ and $232.5 t_0$ respectively. The slope of this linear relationship is about 3 times lower in wet layers compared to dry layers, owing to damping of grain motion by the fluid due the viscous stresses of the fluid phase, consistent with the $\mu(J)$ rheology \cite{pailha2008initiation, rauter2021compressible, fei2023mu, fei2024frictional}. 

\subsection{Fluid Delays Fault Rupture}
\begin{figure}[ht]
\centering
\includegraphics[width=1\textwidth]{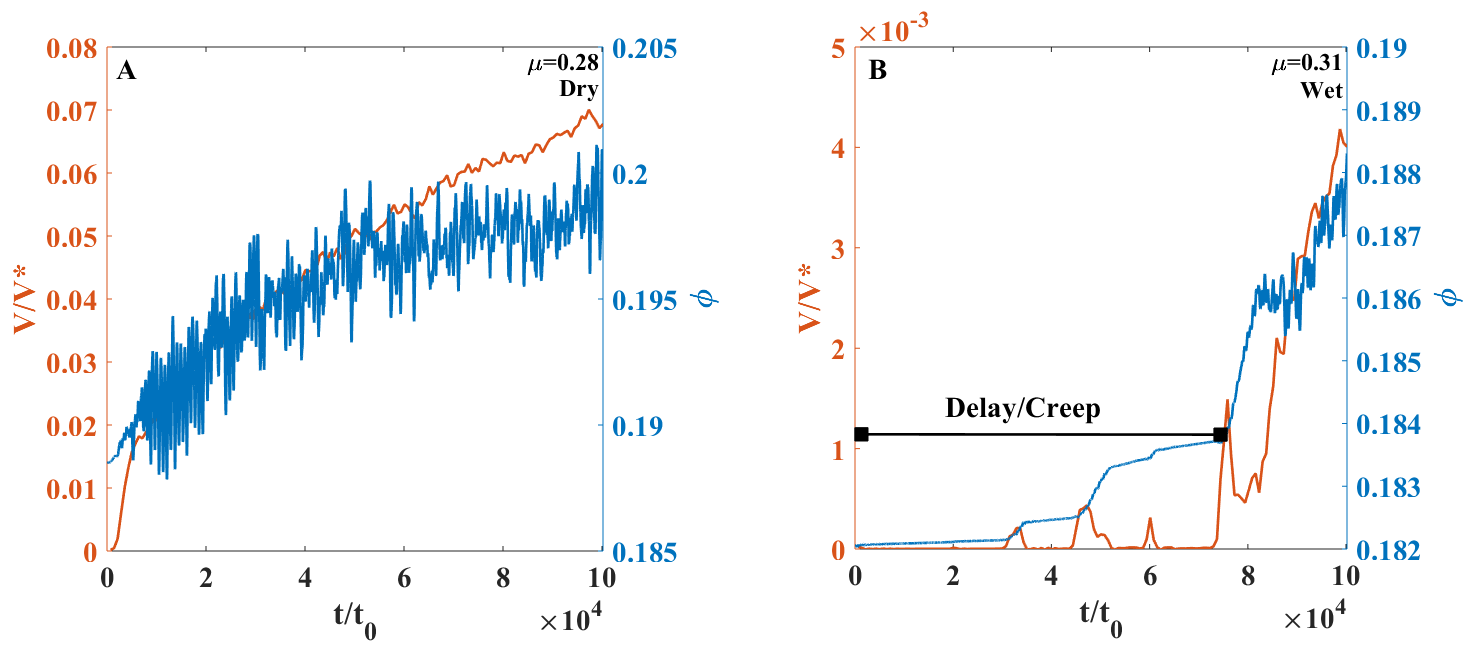}
\caption{Failure of the dry (A) and the wet (B) simulated layer following the application of the critical increment in fluid pressure and shear stress, respectively, at $t=0$. For the wet layer, note the small slip events dilate and weaken the layer until a critical dilation (critical porosity) is reached. This delay is not seen in the dry case.}
\label{fig:fc}
\end{figure}

\begin{figure}[ht]
\centering
\includegraphics[width=1.0\textwidth]{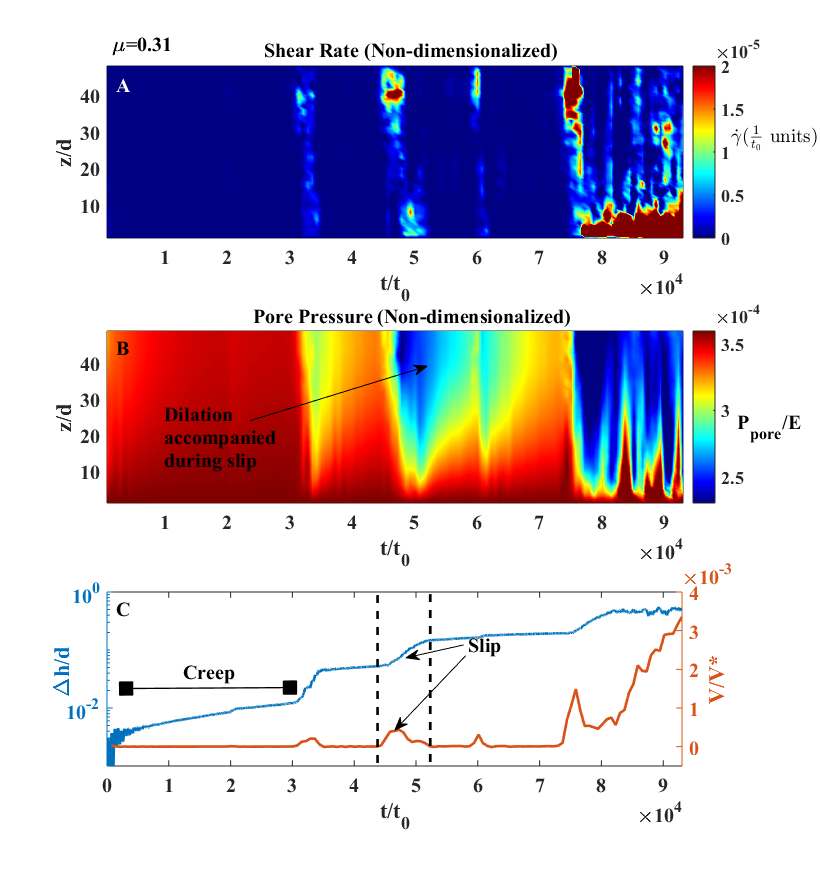}
\caption{Evolution of shear strain rate (A), pore pressure (B), dilatancy, and shear velocity (C) in a critically stressed wet layer following a fluid pressure increment at $t=0$ (from Fig.~\ref{fig:fc}A) . A and B show depth sections. Note the precursory small slip events, accompanied by layer dilatancy and transient pore-pressure drop.}
\label{fig:nass}
\end{figure}

The initiation of motion (branch \#2 in Fig.~\ref{fig:hyster2}) is different in the dry and wet systems. While the dry system fails shortly following the increase in shear stress to its critical level, the wet system shows a delay period between the first application of higher fluid pressure and the onset of large scale slip, as seen in Fig.~\ref{fig:fc}. During this delay time, small slip events occur and are arrested, with each micro-slip accompanied by a small porosity increase. Careful examination of Fig.~\ref{fig:nass} suggests that "dilatancy hardening" slows the onset of rupture \cite{scholz1988mechanisms}, where each dilation event transiently strengthens the layer by transiently reducing the pore-pressure, until diffusion reinstates the pore-pressure to its pre-dilation level. This is seen in Fig.~\ref{fig:nass}, where Panel A and B shows depth sections in time of shear rate and pore pressure just before failure. Panel C shows the corresponding timeseries of dilation and slip velocity. 
Note that two kinds of dilative strain occur prior to failure: a) steady small dilative creep and b) bouts of slip which cause abrupt larger dilation events leading to abrupt drops in pore pressure (Fig ~\ref{fig:nass}B). Pore pressure then relaxes diffusively as the slip  arrests. 

\subsection{Approach to steady state}
Once stress or pressure suffice to initiate large scale motion, the system evolves by approaching a new steady state. Here we analyze the kinematics of the layer and evolution of the state variables during the transient to the new steady state, along branches 2 and 3 in Fig.~\ref{fig:hyster2}, for which large scale motion occurs. 

%Figure \ref{fig:astt} shows evolution of velocity, porosity and coordination number following the failure. Velocity and state variables (porosity and coordination number) have similar transient times for approach to steady state. This is indicative of a strain dependence on friction of the granular layer, because once a particular level of activation strain characterized by the critical porosity is reached, friction drops and the layer accelerates.

We derive an analytical form of the velocity transients in response to step changes in shear stress or injection pressure. We start from the momentum balance

\begin{equation}
    \frac{\mathrm \partial (\rho_s \nu v)}{\mathrm \partial t} = \frac{\partial \tau}{\partial z} \,,
    \label{eqn:mom_bal}
\end{equation}
where $\rho_s$ is the density of grains, $\nu=1-\phi$ is solid fraction, $v$ is (shear) velocity and $\tau$ is the shear stress. The shear stress is proportional to the effective normal stress $\sigma_n'=\sigma_n - P$ via the coefficient of friction: $\tau=\mu\sigma_n'$, where $\mu$ is defined by Eq.~\ref{eqn:stren}. Here we neglect an advection term (the velocity variation along $x$ is negligible due to periodic boundary conditions), and assume $\rho_s$, $\nu$ and $\sigma_n'$ are constant. The latter assumption is only valid if $P_{inj}$ applied at the boundary diffuses sufficiently rapidly across the layer compared to momentum transfer. This assumption is verified for the present setup in Appendix B. Generally, this means that the sample is in a state of equilibrium with the fluid pressure delivered externally, a condition often checked in experimental setups using relatively high-permeability layers \cite{cappa2019stabilization, van2020extracting}. Equilibrium is not the state during the pre-failure micro-slips (Fig.~\ref{fig:nass}), and in fact the lack of equilibrium in those rapid events leads to the dilatancy hardening discussed in the previous section. But once the layer starts sliding fully, the time scale for reaching steady-state momentum balance is longer than that of hydraulic diffusivity (as evidenced by Fig.~\ref{fig:nass}), so that pore-pressure equilibrium may be assumed. In contrast, for shear in low-permeability layers, such as shales, the assumption of constant $\sigma_n'$ breaks down even during the approach to steady sliding, and a relationship between volumetric deformation and fluid pressure must be taken into account \cite{segall1995dilatancy, scuderi2018fluid}.

The momentum balance together with the constitutive law Eq.~\ref{eqn:stren} lead to a diffusion equation for velocity of the grains 
\begin{equation}\label{eqn:diff}
    \frac{\partial v}{\partial t} = D \frac{\partial^2 v}{\partial z^2} \,,
\end{equation}
in which the momentum diffusion coefficient is given by
\begin{equation}\label{eqn:D}
    D = \frac{\sigma_n' \lambda}{\rho_s \nu} \,.
\end{equation}

The solution to Eq.~\ref{eqn:diff} can be sought in a separable form, $v_k=f(z)g(t)$, implying
\begin{equation}\label{eqn:separation}
    v_k(z,t)=[A_k\cos(kz) + B_k \sin(kz)]\exp(-Dk^2 t) \,,
\end{equation}
where constants $k$, $A_k$ and $B_k$ are to be determined from boundary and initial conditions. The full solution for shear velocity is then
\begin{equation}\label{eqn:sol}
    v(z,t)=v_{ss}(z) + \sum_k v_k (z,t)\,,
\end{equation}
in which $v_{ss}$ is the steady-state velocity, obtained as a time-independent solution of Eq.~\ref{eqn:diff}
\begin{equation}\label{eqn:v_ss}
    v_{ss}(z) = G + \chi z\,,
\end{equation}
where $G$ and $\chi$ are constants, and $\chi$ is the steady state shear rate.

We employ the following boundary and initial conditions
\begin{eqnarray}
    &&BC1: v(z=0,t) = 0 \,, \\
    &&BC2: \tau(z=h,t) = \tau_{app} \,, \\
    &&IC: v(z,t=0) = f(z) \,.
\end{eqnarray}
The boundary conditions imply no slip at the bottom surface (BC1) and constant shear stress, $\tau_{app}$, applied at the top surface. The initial condition (IC) specifies the initial velocity profile $f(z)$.

To satisfy BC1, $A_k = 0$ for all $k$, and $G =0$. To satisfy BC2
\begin{equation}\label{eqn:bc2}
    \tau(h) = \sigma'_n \mu(h) = \sigma'_n (\mu_{kinetic} + \lambda \dot{\gamma}(h))=\tau_{app}
\end{equation}
Since Eq.~\ref{eqn:bc2} has to be satisfied for all times, the steady-state shear rate $\chi$ is 
%given by
%\begin{equation}\label{eqn:gammadot_ss}
%    \chi =  \frac{ \tau_{app} - \mu_{kinetic} \sigma_n }{\lambda \sigma_n} \,,
%\end{equation}
determined by
\begin{equation}\label{eqn:nmu2}
  \chi=\frac{1}{\lambda}[\frac{\tau_{app}}{\sigma'_n} - \mu_{kinetic}] 
\end{equation}
and the transient shear rate contributions to stress have to vanish at $z=h$
\begin{equation}\label{eqn:gammadot_transient}
    B_k k \cos(kh) = 0 \,.
\end{equation}
Therefore, permissible constants $k$ are
\begin{equation}
    k_n = \frac{\pi}{2h} + (n-1) \frac{\pi}{h} \,,
\end{equation}
where $n$ is a positive integer.

Taken together, shear velocity is of the form
\begin{equation}\label{eqn:sol_final}
    v(z,t)= \chi z + \sum_{n=1}^\infty B_n \sin(k_n z)\exp(-Dk_n^2 t) \,,
\end{equation}
where the constants $B_n$ are given by the initial condition (IC), $v(z,0)=f(z)$. Specifically, $B_n$ are Fourier series coefficients for the function $f(z)-\chi z$. 

\begin{figure}[htb!]
\centering
\includegraphics[width=1\textwidth]{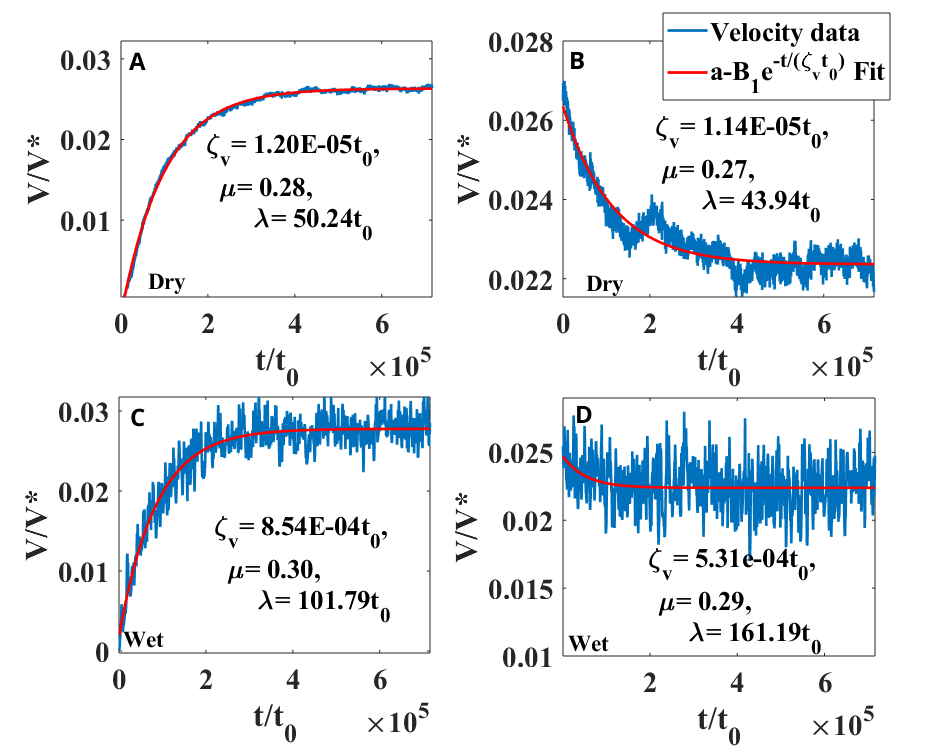}
\caption{Shear velocity evolution (A dry and C wet) during failure (branch 2) and (B dry and D wet) during stress relaxation (branch 3). Numerical data are fit to Eq.~\ref{eqn:vt} (at $z=h$), in which the transient time $\zeta_v$ (in units of $t_0$) is considered as a free parameter. The value of the strengthening coefficient $\lambda$ (in units of $t_0$)is inverted from the transient time using Eq.~\ref{eqn:transient_time}. }
\label{fig:fit2}
\end{figure}

Note that the transient terms decay rapidly with increasing $n$. For sufficiently large times, velocity is thus dominated by the steady-state term and $n=1$ term, which has the slowest time decay,
%We should compare this decay with the velocity transients of the form 
\begin{equation}\label{eqn:vt}
    v(z,t) \approx \chi z + B_1 \sin \left(\frac{\pi}{2h} z \right) \exp\left(- \frac{t}{\zeta_v} \right) \,.
\end{equation}
Where $\chi$ is defined in Eq.~\ref{eqn:nmu2} and the transient time $\zeta_v$ is inversely proportional to the strengthening coefficient $\lambda$, following:
\begin{equation}\label{eqn:transient_time}
    \zeta_v = \frac{4}{\pi^2} \frac{\rho_s \nu h^2}{(\sigma_n-P_{inj}) \lambda } \,,
\end{equation}
 Both $B_1$ and $\lambda$ are fitting parameters, obtained by fitting Eq.~\ref{eqn:vt} to velocity transients from our simulations (see Fig.~\ref{fig:fit2}). The steady-state velocity, $v_{ss}= \chi z$, is, based on Eq.~\ref{eqn:nmu2}
\begin{equation}\label{eqn:v_p_ss}
    v_{ss}(z)=[\frac{ \tau_{app}}{ (\sigma_n-P_{inj})}-\mu_{kinetic} ]\frac{z}{\lambda}\,.
\end{equation}

 Fig.~\ref{fig:fit2} tests the analytical prediction by comparing the simulated top wall  velocity to Eq.~\ref{eqn:vt}, with $V=v_{ss}(z=h)$, where time, $\zeta_v$ and $\lambda$ are in units of $t_0$, as noted in Table \ref{tab:normal}. The transient slip velocity corresponding to shear stress or fluid pressure increase (Fig.~\ref{fig:fit2} A \& C) or decrease (Fig.~\ref{fig:fit2} B \& D) can indeed be exponentially fitted. The strengthening coefficient $\lambda$ resulting from the best fit is similar for an increase in loading (acceleration on branch 2) as well as a decrease (deceleration on branch 3), although the data for the decrease are more noisy and less sensitive to the value of the fitting parameter. It is also consistent within a factor of 2 with the steady-state calibration obtained in Fig.~\ref{fig:fit1}. The discrepancy between the steady-state and transient calibration, apparent particularly for the wet layer, is likely because of pore pressure variation during the transient phase, as evident in Fig.~\ref{fig:nass}.

\begin{figure}[htb!]
\centering
\includegraphics[width=1.1\textwidth]{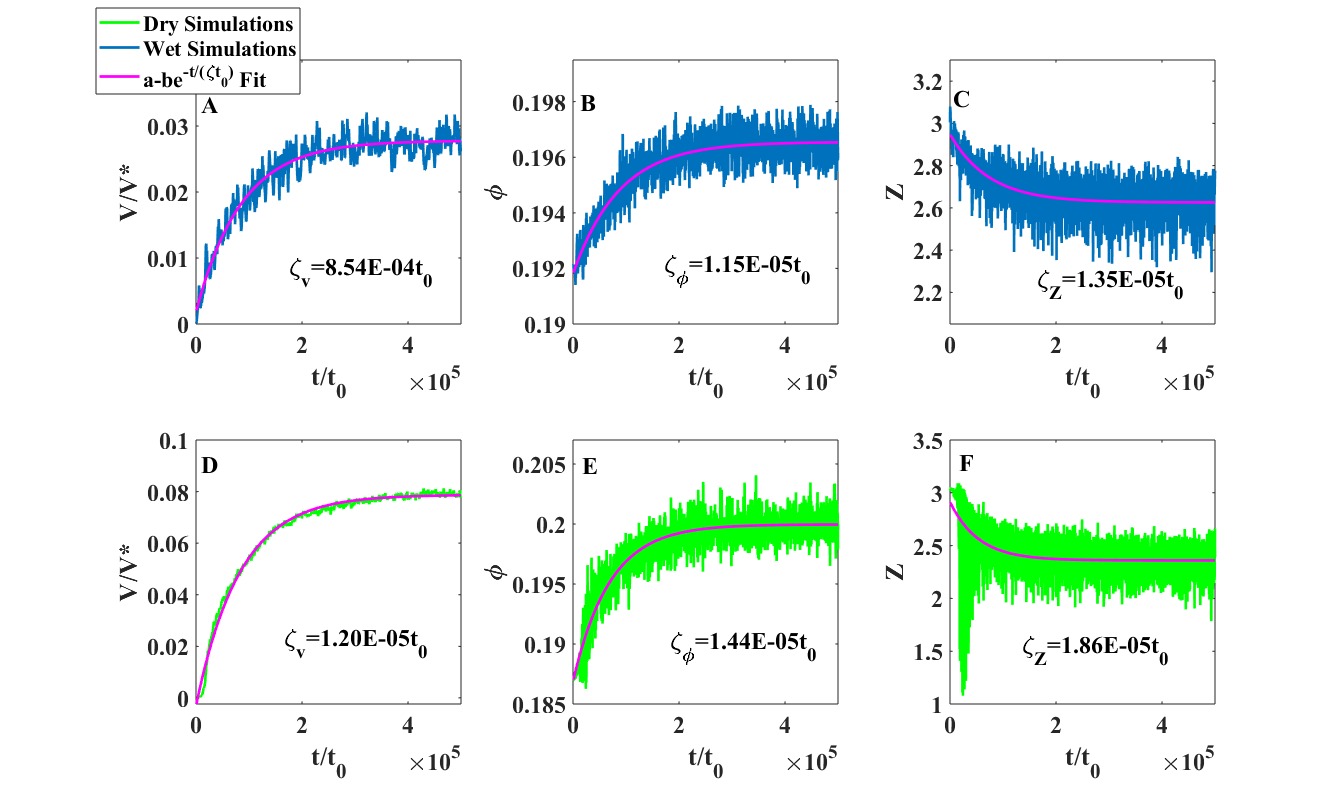}
\caption{Approach to steady state in the accelerating layer after failure (branch 2), for dry(green) and wet(blue) simulations showing velocity (A \& D), porosity (B \& E), and coordination number (C \& F). Note the similar transient time $\zeta$ (in units of $t_0$)for the three quantities under both dry (green) and wet (blue) conditions.}
\label{fig:astt}
\end{figure}

Finally, Fig.~\ref{fig:astt} demonstrates that porosity and coordination number show a similar transient time $\zeta_v$ for their approach to steady state. Hence, the evolution of state variables is tightly related to the evolution of pore-pressure within the layer.

\section{Discussion}
This work studies initiation and cessation of shear in fluid saturated granular layers driven by fluid pressure, providing three main results: 1) a hysteresis loop of friction, porosity and grain co-ordination number. 2) velocity-strengthening friction, and 3) a delay in slip onset following pressure injection, which does not appear in dry (control) systems.  

\subsection{Characteristics of Hysteresis}
The hysteresis loop in fluid-filled systems is similar to that in dry systems, and is characterized by three unique modes of motion: first, as the shear stress or injection pressure are increased, but remain below the failure limit, the system responds by  elastic ringing  interspersed  with micro-slips (Fig.~ \ref{fig:hysterrep}A \& D). Once the critical failure limit is reached, the layer accelerates to steady state with a characteristic transient time (Eq.~\ref{eqn:transient_time} and Fig.~\ref{fig:hysterrep}B \& E), controlled by momentum diffusion (Eq.~\ref{eqn:diff}). Once in motion though, if the stress or pore pressure are reduced, the granular system  doesn't retrace its path to stagnation. Instead,  reduction of shear stress or injection pressure decelerates the motion with a linear dependence between slip velocity and shear stress, until the layer halts. The loading loop shows hysteresis manifested by higher friction coefficient $\mu_{static}$ required for initiation of slip than that observed when slip ceases, $\mu_{kinetic}$. This hysteretic behavior has been observed for dry GM previously, both experimentally and in simulations \cite{nasuno1997friction,daniels2005hysteresis, garcia2008strong, degiuli2017friction}. Our work extends this behavior for fluid-filled GM subject to fluid injection. State variables, namely porosity and grain coordination number, also exhibit similar hysteresis. 
%,perrin2019interparticle
In addition, we explore how the amplitude of the hysteresis $\Delta\mu=\mu_{static}-\mu_{kinetic}$ varies with layer aspect ratio. $\Delta\mu$ is found to decrease with aspect ratio, until $l/h \approx 1.5$, after which it saturates, possibly due to the number of force chains in the system, relative to their length. Another important constraint on the amplitude of hysteresis is the dimensionality of the simulations. Since our model is essentially two dimensional, the hysteresis amplitude is expected to be larger than that in the full 3D case \cite{frye2002effect}. However, hysteresis is expected to be  maintained in 3D \cite{nasuno1997friction, frye2002effect, daniels2005hysteresis, garcia2008strong, scholtes2012discrete, martin2020micromechanical, hong2022relaxation}, since it is a reflection of the transition from static to dynamic friction, irrespective of the dimensionality. 

The mechanical origin of the hysteresis in GM is a subject of ongoing research \cite{mowlavi2021interplay}. Some of the proposed mechanisms include the role of interparticle friction \cite{degiuli2017friction, perrin2019interparticle} and dissipation of energy which was stored elastically within the grains during loading \cite{radjai1998bimodal}.
However, probably the geometry, particle shape, packing structure and most of all porosity of the granular layer, play a crucial role in the hysteresis under cyclic loading. Shear resistance depends strongly on porosity \cite{campbell2006granular, chen2016rate, amiri2019development}, so a compact layer is harder to shear than a loosely packed layer. The fact that GM must dilate in order to shear has already been well established more than a  century ago  \cite{reynolds1885lvii, reynolds1886experiments}. Once  the system is dilated to its critical level, and motion initiates, the shear resistance drops from $\mu_{static}$ to $\mu_{kinetic}$. Additionally, the arrangement and packing structure may be  directly linked to the grain coordination number $Z$, ordered packing has a higher value of $Z$ and hence tends to be more stable, with higher frictional resistance, compared to random packing \cite{tordesillas2007force, dijksman2011jamming}. In natural or more complex systems, additional controls on hysteresis may arise from a) grain angularity as irregularly shaped particles that interlock more effectively, increasing the frictional resistance due to increased energy dissipation during particle rearrangements, relative to spheres that roll and slide more easily  \cite{mair2002influence, nouguier2003influence, guo2004influence, anthony2005influence, santamarina2009friction, jerves2016micro} and b) breaking and pulverization of grains during shear which lead to localization \cite{abe2009effects, barras2023shear}.

\subsection{Competition between dilation and pressure}

During the approach to ultimate failure, the resisting shear traction in preparation to unstable motion of the layer is controlled by two factors, a) the effect of pore pressure, and b) the effect of porosity \cite{reynolds1885lvii, reynolds1886experiments,marone1990frictional}. The first effect is known as "dilatancy hardening", which simply put says that as the system dilates, pore pressure drops and shear resistance of the layer increases \cite{scholz1973earthquake, segall1995dilatancy}.  
The second effect, discovered by \citeA{reynolds1885lvii,reynolds1886experiments}, is due to highly porous GM being weaker, less resistant to shear, than densely packed GM. The fact that the shear strength of GM (both dry and wet), and also that of intact porous matter, decreases with increasing dilation (porosity)  has been noted widely \cite{kruyt2006shear, campbell2006granular, niemeijer2007microphysical}. In addition, the dilation process works against the normal force and requires energy which translates into a friction increase \cite{makedonska2011friction,marone1991note}. 

We dont attempt to write here a full descriptive and coupled model of shear resistance as function of pore pressure and dilation. But a very simple and illustrative model of shear strength as function of pore-pressure and porosity may look like this:
\begin{equation}
    \tau=\mu(\phi) (\sigma_n-P)=\frac{\mu_s^c+2H(\phi_c-\phi)}{1-2\mu_s^cH(\phi_c-\phi)}(\sigma_n-P)
    \label{eqn:chen}
\end{equation}
where the strength is a product of two terms that evolve with shear displacement: first the friction coefficient, $\mu(\phi)$, and second the pore-pressure,  both depending on porosity. 
Since the approach to failure is characterized by ongoing dilation (as seen in Fig.~\ref{fig:nass}, and also in experiments e.g. \citeA{guglielmi2015seismicity, cappa2019stabilization}), both terms evolve as failure approaches. 
The functional dependence of $\mu$, the apparent friction coefficient, on porosity $\phi$, can be easily calculated analytically if we assume spherical uniform grains. Eq.~\ref{eqn:chen} uses the analytical grain model derived by \citeA{chen2016rate}, where $\mu_s^c$ is the static friction of the grain contacts, $H$ is a geometric constant and $\phi_c$ is the critical state porosity for the granular flow.  Eq.~\ref{eqn:chen} predicts that  $\frac{d \mu(\phi)}{d\phi}<0$, so that as porosity increases, and the grains climb on each other in preperation for sliding, the friction coefficient drops. Yet if porosity increases rapidly enough, the pressure simultaneously drops (see Fig.~\ref{fig:nass}). This makes the entire approach to failure a competition between dilation and pore pressure. Dilation hardening repeatedly arrests slip bouts, by pore pressure reduction, as seen Fig.~\ref{fig:nass} and predicted theoretically in \citeA{segall1995dilatancy}. But as porosity slowly increases, with each bout of slip the layer strength, $\mu(\phi)$, is  degraded, until the layer finally fully fails, since the strength, which is the product of the two terms in Eq.~ \ref{eqn:chen} drops below the applied stress. 

The competition between $\phi$ and $P$, identified  by Eq. ~\ref{eqn:chen}, means there is no unique porosity or pressure value for which the system fails. Instead, for each porosity value the system attains, there will be a high enough pore pressure that can fail this layer, as can be seen in Eq.~\ref{eqn:chen}. Conversely, if the pressure is a bit low, the system may dilate more in order to fail. This interplay is further complicated via  pressure reduction during dilation, and restoration by diffusion. Thus the GM acts as a dynamical system that reaches failure along an optimal $P-\phi$ line.

\subsection{Delay in sliding onset}
\begin{figure}[ht]
\centering
\includegraphics[width=1\textwidth]{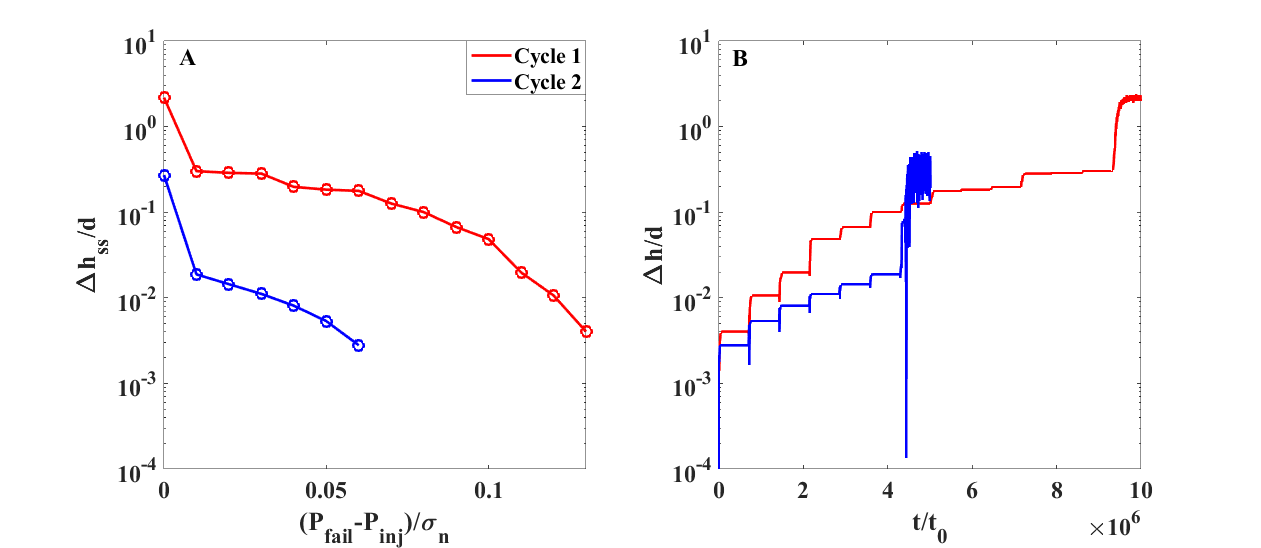}
\caption{Dilation of the layer in wet case for the first two loading cycles with respect to the non-dimensional pressure difference from failure $\frac{P_{fail}-P_{inj}}{\sigma_n}$ can be seen in A. In B this same dilation is shown as a time series, from the beginning of injection till it approaches failure.}
\label{fig:dilts}
\end{figure}

The initiation of motion in fluid-filled sheared GM is characterized by a delay, related to an initial permanent porosity increase and exclusively only observed in the first pressure cycle, and for fluid-filled systems.
Similar delay in onset of catastrophic failure has been observed experimentally for underwater granular avalanches, and was linked to the degree of initial compaction of the grains \cite{pailha2008initiation}. The source of this delay has been later theoretically elucidated by coupling the dilatancy of the granular layer with the interstitial fluid pressure and suggesting a sharp sensitivity of the delay time to the initial volume fraction of the granular layer (dilatancy angle) \cite{pailha2009two}. 
In context of fluid-filled fault zones, this observed delay is consistent with previous theoretical studies of fluid-filled granular fault gouge \cite{van2020extracting} and with actual field experiments of fluid injection into fault zones \cite{guglielmi2015seismicity}. Fig.~\ref{fig:dilts} shows this preparatory dilation (prior to failure) as a function of time (panel B) and distance from failure in terms of $(P_{fail} - P_{inj})/\sigma_n$ (panel A), where $P_{fail}$ is the value of fluid pressure at the boundary at which the layer fails. As we approach failure, the layer accumulates dilation, up to a threshold which is close to $\frac{1}{10}$th of the grain diameter, after which the layer yields to the applied traction. 
The dilatant creep occurs mostly via a series of attempted sliding events, which competes with the rate of pore-pressure diffusion. Each slip events is seen to reduce pore-pressure (Fig.~\ref{fig:nass}), and is followed by a relatively long wait time until the next event. Since dilatancy predominantly occurs during these abrupt and disperse events, this halted process slows down the onset of large scale failure. This process agrees with the proposal that slowing of rupture may be produced by dilatant hardening \cite{segall2010dilatant, ikari2013slip, chen2023emergence}. However, dilation is  necessary  for slip to occur in GM, as already noted by \citeA{reynolds1885lvii}. This is a configurational issue that has nothing to do with the fluid, porous granular layers are in general easier to shear than compact ones. Only after the layer reaches its "critical porosity", can it shear, and  rupture will occur. The time it takes the system to reach this critical porosity controls the rupture onset time.  
However the details of the slow slip onset is seen here to be different than previously suggested - instead of a slow monotonic rise in dilatancy opposed by a slow infiltration of pore-pressure, we see a series of attempted, but quickly halted, slip events. In this way dilatancy hardening stabilizes slip (as proposed in \citeA{rudnicki1988stabilization, williams2024effects}), yet eventually after a series of such events, finally accelerating rupture occurs.
The time scale of the delay is controlled by the rate of dilation, which occurs both via creep and via bursts, interspersed (Fig.~\ref{fig:nass}). It is also dictated by the pore pressure level, and its equalization rate, which in turn is controlled by hydraulic diffusivity.

\subsection{Shear-rate strengthening}
In the inertial regime, after the layer starts accelerating and sliding fully, strain-rate dependence of friction governs the transient behavior theoretically predicted in section \ref{srt}. 
The shear strength of our granular layer increases with increasing strain-rate (velocity) following Eq.~\ref{eqn:stren}. As the strain-rate increases, grain inertia and energy dissipated in collisions increase, leading to an increase in the friction coefficient. This behavior in GM is described by a $\mu(I)$ rheology \cite{forterre2008flows, GDRMiDi, daCruz2005, da2002viscosity} for dry GM and by a $\mu(J)$ rheology \cite{pailha2008initiation, rauter2021compressible, fei2023mu, fei2024frictional} for wet GM when viscous stresses of the fluid phase also become important. In both of these cases friction depends linearly on strain-rate. This strain-rate strengthening produces steady-state motion at rates dictated by the strengthening coefficient and applied shear stress. For a fluid-filled layer, the level of pore pressure will dictate steady state slip rate, as reflected in our theory (Eq.~\ref{eqn:vt} and $v_{ss}$ in Eq.~\ref{eqn:v_p_ss}). Strengthening via dynamic dissipation willl occur in fast moving faults, moving at close to seismic slip rates \cite{rubino2022intermittent}. At slow shear rates, experimental observations on bare surfaces measure logarithmic dependence of steady-state friction on shear-rate, but at higher slip velocities this dependence  transitions to stronger-than-logarithmic shear-rate strengthening, that is typically accompanied by a change in the dominant frictional dissipation mechanism \cite{bar2014velocity}. 

\subsection{Implications to natural systems}
Most of our implications relate to onset and cessation of slip in fluid-filled faults: 
First, fluid-saturated granular faults exhibit episodic small slip events reflecting arrested ruptures by dilation hardening, which precede full-blown rupture. These episodic slips serve as a "warning sign" that the fault is dilating in preparation for a future accelerating rupture. The time-scale of this preparatory phase is not predictable at this stage, yet it relates to the rate of porosity increase and the coupling between dilation rate and pore pressure evolution, in total providing the evolution of the fault strength, as  predicted for a simplified setting by Eq.~\ref{eqn:chen}. Similar dilative motion-bursts prior to a large earthquake have been observed to precede the 2004 Parkland Earthquake on the San-Andreas Fault \cite{khoshmanesh2018episodic} and are similarly correlated to pore pressure fluctuations, although their origin has not been previously explained. 

Second, in the slip-preparation phase we observe, similarly to \citeA{proctor2020direct}, that the pressure fluctuations during dilatant slip bursts (Fig.~\ref{fig:nass}B) are large, up to 20-30\% of the normal stress. This means that stress fluctuations due to variations in effective normal stress vary fault strength much more significantly than rate and state frictional effects, since the latter are predicted to change fault-strength by only a few percent  \cite{aharonov2018physics}, and are not included in this study. If similar pore-pressure variations during the slip onset carry on to larger 3D systems (and this should be studied), one important implication is that strain-weakening and strengthening effects, due to pore-pressure fluctuations and porosity changes, may overshadow strain-rate effects in controlling the path of fluid-saturated faults to failure.  

A third major implication concerns the observation that GM shearing follows a velocity- strengthening law (as predicted by Eq.~\ref{eqn:stren}). This result agrees with the generally accepted strain-rate dependent rheologies, whether it is the $\mu(I)$  for dry GM \cite{da2002viscosity, GDRMiDi, daCruz2005, forterre2008flows} or $\mu(J)$ for fluid-filled GM when viscous stresses become important \cite{pailha2008initiation, rauter2021compressible, fei2023mu, fei2024frictional}, and additionally is also observed experimentally \cite{rubino2022intermittent}. In the field, faults may also show velocity-strengthening sliding, as indicated by the observation of aseismic slip induced by fluid-injection \cite{cornet1997seismic, guglielmi2015seismicity, cornet2016seismic}, yet the mechanism of this velocity- strengthening may be different than the dynamic dissipation which leads to the $\mu(I)$ rheology. Our work finds, within the velocity strengthening branch, a correlation between fluid pressure level and slip rates, which can be applied to other cases of velocity strengthening, even though the strengthening mechanism may be different (e.g. \citeA{perfettini2008dynamics, bar2014velocity}). Additionally, such relationships between slip rate and the traction on the fault, specifically due to the presence of fluids, have been experimentally observed \cite{scuderi2017frictional}.

The fourth implication arises from the observed hysteresis. In particular, consider the following fluid injection scenario: A fault with a background pressure $P_0$, started sliding following an injected pressure increase $\Delta P$. Removing $\Delta P$ and returning to $P_0$, will not stop the slip. Because of the hysteresis, once a fault starts slipping due to injection it will continue to slip even after pore-pressure is reduced to its original value, $P_0$, as it will be within the sliding  (\#3) branch of the hysteresis in our work (Fig.~\ref{fig:hyster2}). This can explain why earthquakes still occur post shut-in of injection in major field fluid injection cases \cite{healy1968denver, deichmann2009earthquakes, bachmann2011statistical, ellsworth2019triggering, saez2023post}.

%Text here ===>>>

\section{Conclusions}
\begin{enumerate}
    \item Frictional hysteresis in dry GM has been widely studied, here we extend this study to sheared GM subjected to fluid injection. We highlight the major differences and similarities between both cases. The discovery of a  frictional hysteresis in fluid-injected GM, suggests that simply reducing fluid pressure just below the failure pressure predicted by the Mohr-Coulomb failure criterion, is insufficient to stop a granular fault gouge which is already  sliding, and further reduction of pressure  below the failure pressure is necessary to arrest slip. This has implications on post shut-down seismicity of fluid injection in field cases.
    \item Dry and wet GM differ in their approach to  failure. Where a dry granular layer fails immediately once enough shear stress is applied on the layer, fluid injected granular layers show a delay. This delay is characterized by bouts of small slip events which are arrested owing to dilatant hardening, yet serve to dilate the layer and thus weaken it, until the layer reaches a critical porosity after which it undergoes ultimate failure. 
    \item Large pore-pressure drops accompanied by episodic slip in preparation for ultimate failure, suggest that for fluid-saturated faults, stress fluctuations due variations in effective normal stress may be more dominant over rate and state effects, in controlling the path to failure. 
    \item We theoretically predict velocity transients and approach to steady-state sliding,  using a constitutive shear-rate strengthening friction relationship. This connects slip rates of granular fault to the level of fluid pressure injected into the system, suggesting that for granular fault gouge higher injected pressure will lead to higher slip rates.

\end{enumerate}

\section*{Open Research Section}
The simulation software for this research can be found at \citeA{sarma_2024_13765085}, available via CC-BY-NC license. Raw numerical data were post-processed and figures were made with MATLAB software version R2023b, available under the academic license at www.mathworks.com. The MATLAB script used for the post-processing is available along with the simulation codes at \citeA{sarma_2024_13765085}. Datasets for results of simulations are available at \citeA{sarma_2024_13764960}.

\acknowledgments
PS and EA would like to acknowledge the support of the Bi-national Israel-US Industrial Development fund of the US-Israel Energy Center for Fossil Energy. RT acknowledges the support of the University of Oslo, the Njord Centre, the CNRS IRP D-FFRACT and the Research Council of Norway through the PoreLab Center of Excellence (project number 262644). SP acknowledges the support of the Internal Grant Agency of Jan Evangelista Purkyne University (project UJEP-IGA-2024-53-003-2).

\appendix
\section{Coupled Solid-Fluid DEM formulation}\label{AppendixA} 

\subsection{Fluid Phase} 
In the works \cite{goren2011mechanical,goren2010pore, niebling2010sedimentation, niebling2010mixing, niebling2012dynamic}, the mechanics of the pore fluid coupled with granular motion was developed in detail. We'll go through it briefly here for clarity. First, mass-conservation equations for grains and fluids are considered.
\begin{equation}
\label{mass_s}
    \frac{\partial[(1-\phi)\rho_s]}{\partial t} + \nabla \cdot [(1-\phi)\rho_s \mathbf{u_s}]=0 \,,
\end{equation}
\begin{equation}
\label{mass_f}
   \frac{\partial[\phi \rho_f]}{\partial t} + \nabla \cdot [\phi \rho_f \mathbf{u_f}]=0 \,,
\end{equation}
where $\rho_s$ and $\rho_f$ are the densities of the solid grains and fluid, respectively, $\mathbf{u_s}$ and $\mathbf{u_f}$ are the solid and fluid velocity fields, respectively, $\phi$ is porosity and $t$ is time. These velocities are defined for mesoscopic volumes of at least a few grains. 

Fluid momentum equation is here approximated by Darcy's law:
\begin{equation}
\label{Darcy}
 \phi(u_f-u_s)=-\frac{k}{\eta}\nabla P \,.
\end{equation}
Here $P$ is the excess fluid pressure, above the hydrostatic level. This formulation neglects fluid inertia as well as viscous drag forces.
%In this equation, $\eta$ is the fluid viscosity, $P$ is the excess fluid pressure (above hydrostatic), and $k$ is the permeability. 
Next, we consider a linearized fluid state equation
\begin{equation}
\label{state_f}
\rho_f=\rho_0(1+\beta P) \,,
\end{equation}
where $\beta$ is adiabatic fluid compressibility and $\rho_0$ is the density of the fluid at the reference hydrostatic pressure. For the solid phase, we assume its compressibility is much lower than the fluid compressibility, and therefore $\rho_s$ is constant.

Combining Eqs. \ref{mass_s}-\ref{state_f} leads to the following governing equation for excess pore pressure \cite{goren2011mechanical,Parez2023b}
\begin{equation}
\frac{\partial P}{\partial t} - \frac{1}{\beta \phi \eta} \nabla \cdot [k \nabla P] + \frac{1}{\beta \phi} \nabla \cdot \mathbf{u}_s = 0 \,.
\end{equation}
This equation is solved on a regular grid with spacing equal to two mean grain diameters. The properties of the porous granular matrix, $k$, $\phi$ and $\mathbf{u}_s$, may vary in space and are interpolated from grains onto the grid points using the bilinear interpolation scheme \cite{goren2011mechanical}.

Permeability and porosity in the model are related by Carman-Kozeny relationship. Since we use 2D geometry, we map our 2D porosity to a 3D field assuming random close packing \cite{mcnamara2000grains} to arrive at the following constitutive law for permeability
\begin{equation}
k=\frac{k_c(1+2\phi_{2D})^2}{(1–\phi_{2D})^2}
\end{equation}
where $k_c$ is a prefactor used to scale permeability to a desired order of magnitude and $\phi_{2D}$ is the 2D porosity.

\subsection{Solid Phase} 
A discrete element model with linear elastic frictional contact model is used to solve grain dynamics \cite{Cundall1979}. In DEM, grains are represented as spheres interacting via contact forces. The pair-wise contact force includes elastic repulsion, velocity-dependent damping and a Coulomb friction force \cite{parez2021strain,Parez2023b}. The linear and rotational momentum conservation equations 
\begin{equation}\label{eqn:tr}
m_i \dot{\mathbf{u}}_{s,i}=\sum_j \mathbf{F}_{ij} - \frac{V_i}{1-\phi}\nabla P
\end{equation}
\begin{equation}\label{eqn:rt}
I_i \dot{\bm{\omega}}_{s,i}= \sum_j R_i \hat{\mathbf{n}}_{ij} \times \mathbf{F}_{ij}
\end{equation}
are time integrated using the Verlet algorithm.
Here, $\dot{\mathbf{u}}_{s,i}$ and $\dot{\bm \omega}_{s,i}$ are the translational and rotational accelerations of grain $i$, and $m_i$ and $I_i$ are its mass and moment of inertia. For the linear momentum, the first term on the right-hand side is the sum of contact forces $\mathbf{F}_{ij}$ with all grains $j$ that are in contact with grain $i$. The second term is the drag force proportional to pore pressure gradient $\nabla P$ and volume of the grain, $V_i$. In the equation for rotational momentum, $R_i$ is the radius of grain $i$ and $\hat{n}_{ij}$ is a unit vector along the direction connecting the centers of grains $i$ and $j$.

\bibliography{library}

\end{document}